\documentstyle[aps,prl,multicol,fancyheadings,epsfig,amsbsy,amssymb,amstex]{revtex}

\def\bbbc{{\mathchoice {\setbox0=\hbox{$\displaystyle\rm C$}\hbox{\hbox
to0pt{\kern0.4\wd0\vrule height0.9\ht0\hss}\box0}}
{\setbox0=\hbox{$\textstyle\rm C$}\hbox{\hbox
to0pt{\kern0.4\wd0\vrule height0.9\ht0\hss}\box0}}
{\setbox0=\hbox{$\scriptstyle\rm C$}\hbox{\hbox
to0pt{\kern0.4\wd0\vrule height0.9\ht0\hss}\box0}}
{\setbox0=\hbox{$\scriptscriptstyle\rm C$}\hbox{\hbox
to0pt{\kern0.4\wd0\vrule height0.9\ht0\hss}\box0}}}}
\newcommand{\beq}{\begin{eqnarray}} 
\newcommand{\eeq}{\end{eqnarray}} 

\begin{document}
\title{Visualizing the particle-hole dualism   in   high temperature
   superconductors.}
\author{I. Martin and A.V. Balatsky}
\address{Theoretical Division, Los Alamos National Laboratory, Los Alamos, NM 87545}

\date{Printed \today }

\maketitle

\begin{abstract}
Recent Scanning Tunneling Microscope (STM) experiments offer a unique 
insight into the 
inner workings of the superconducting state of high-Tc superconductors.
Deliberately placed inside the material impurities perturb
the coherent state and produce 
additional excitations.  Superconducting excitations --- quasiparticles ---
 are the quantum mechanical mixture of negatively
charged electron (--e)  and positively charged hole (+e). 
Depending on the applied
voltage bias in STM one can sample the particle and hole content of 
a superconducting excitation.  
We argue that the complimentary cross-shaped patterns observed 
on the positive and negative biases are the manifestation of the 
particle-hole dualism of the quasiparticles.

\end{abstract}
\pacs{Pacs Numbers: XXXXXXXXX}

\vspace*{-0.4cm}
\begin{multicols}{2}

\columnseprule 0pt

\narrowtext
\vspace*{-0.5cm}

The dual particle-wave character of microscopic objects is one of the 
most striking phenomena in nature.  
While posing deep philosophical problems, the dualism is ubiquitous 
in the microworld.  Most notably, the two-slit experiments of Stern and 
Gerlach revealed the interference and, hence, the wave nature of 
electrons.  In the condensed matter systems, such explicit visualization
of the wave nature of the constituent electrons was missing until just 
recently.  The breakthrough came when the researchers from the IBM labs 
realized that the best way to elucidate the electrons {\em inside}
a material is to place an impurity in an otherwise perfect 
crystal structure.
By building corrals of the impurities on the clean surface, and observing the 
generated patters through the scanning tunneling microscope  (STM),
the experimenters were able to 
demonstrate the laws of the wave optics using the conduction 
electron waves.\cite{coral}

The analog of the conduction electrons in the superconductors are the 
quasiparticles.  Unlike electrons, the superconducting quasiparticles do not
carry definite charge.  Like the Cheshire cat, the quasiparticle is  
a combination
of an electron and its absence (``hole'').  And much like the Cheshire cat,
the superconducting quasiparticles have never been seen in nature.
Until now.  In the series of beautiful experiments 
J.C. Davis' group \cite{Pan1,Pan2}
observe just that -- the interference of the superconducting quasiparticles,
which depending on the way one looks at them show their electron or hole
parts. 

Pan {\em et al.} explore the structure of the superconducting
state in Bi$_2$Sr$_2$CuO$_2$ high-temperature superconductor in the 
vicinity of Ni and 
Zn impurities.
To visualize the local quasiparticle states they employ the STM technique.
There is one aspect of the electron tunneling into
the superconducting state that makes it qualitatively different
from the tunneling in conventional metals. 
The STM tip contains only the regular electrons which carry a unit of 
charge (--e).
On the other hand, quasiparticles that live inside the superconductor 
do not possess a well-defined charge.
Upon entering the superconductor, an electron that arrived from the
normal STM tip must undergo a transformation into the Bogoliubov quasiparticles
native to the superconductor.\cite{Schrieff}  The detailed 
process of conversion of
an electron into a quasiparticle is a deep  
theoretical problem. 
  Another example of such process occurs in the 
case of the tunneling into fractional quantum Hall liquid, where the natural 
quasiparticles carry a fractional, but well defined, charge.
Fortunately, for the low intensity tunneling exact details of the 
particle-quasiparticle conversion become irrelevant, and the tunneling 
amplitude is simply determined by the overlap between the electron state and
the quasiparticle state of the same energy.  That is to say that the tunneling
intensity is high if the ``electronic content'' of the quasiparticle is high, 
and wise versa.

The most striking experimental observation that Pan {\em et al.} make is 
that close to the impurity additional electronic states are generated, with
the energies inside the superconducting gap.  That such states should
exist in conventional (s-wave) superconductors was first predicted by 
Shiba and  others \cite{shiba,yulu} in the late 1960's, while for the unconventional 
(d-wave) 
superconductors these states were predicted and intensively studied in
\cite{balatsky,byers,flattereview}. Experimentally, these intra-gap 
states in conventional (Nb)
superconductor were previously observed in IBM experiments \cite{Yazdani}. 
 
The low-lying impurity states are produced when the local 
impurity is sufficiently 
strong so as to significantly disturb the superconducting order 
parameter in its neighborhood.
It has been found theoretically that in the conventional superconductors an impurity that 
doesn't have it's own magnetic
moment, or ``spin,'' should not produce states inside the gap.
On the other hand, in the d-wave superconductors, 
even a potential (spinless) impurity produces two states located 
symmetrically above and below the chemical potential.

In the latest experiments, Pan {\em et al.} see the impurity states 
inside the gap.  However,
the most surprising is the spatial structure of the impurity states.  
For a Ni impurity, the 
positive-energy state, which corresponds to adding an electron, has the largest
weight at the impurity site, smaller weight on the next-nearest neighbors and 
even smaller weight on the nearest neighbors.
The negative-energy state, which corresponds to removing an electron, shows a 
complimentary pattern: The impurity state is evenly distributed between the 
impurity's four nearest neighbors, with almost no weight at the impurity 
site, or at the next-nearest neighbors.  

In this paper we show how this 
highly non-trivial impurity state structure can be accounted 
for by using the particle-hole 
dualism of quasiparticles in superconducting state.  We implement
simple 
but realistic model of the doped cuprate superconductor with potential
impurity.  We demonstrate that the alternating intensity of the impurity
states is the manifestation of the quantum wave nature of the quasiparticles
scattering from the impurity.

\begin{figure}[htbp]
  \begin{center}
    \includegraphics[angle = -90, width = 3.0 in]{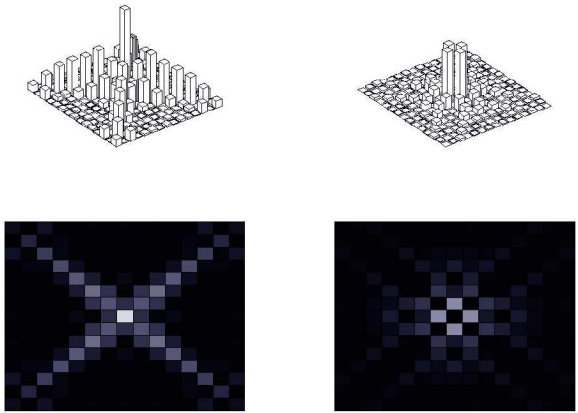}
\vspace{0.5cm}

\caption{Impurity state patterns for a positive (left) and negative (right) bias.
Impurity strength is $V_{imp} = -3t$, doping $14\%$.}
\label{fig:best}
\end{center}
\end{figure}

Qualitatively, the spatial distribution of tunneling intensity 
can be understood as follows.
Let us define the respective 
amplitudes of particle and hole parts of the Bogoliubov
quasiparticle, $u_n(i)$ and $v_n(i)$ for site $i$ and 
for particular eigenstate $n$.  They obey the normalization 
condition $\sum_n |u_n(i)|^2 + |v_n(i)|^2 = 1$ for any 
fixed site $i$. Consider now a site where, say, $u_n(i)$ 
is large and close to 1. It follows therefore that for 
the same site the $v_n(i)$ would have to be small, 
since the normalization condition is almost fulfilled
by $|u_n(i)|^2$ term alone. Similarly, for the sites
where $v_n(i)$ has large magnitude, $u_n(i)$ would 
have to be small. Recall now that large $u(i)$ 
component would mean that quasiparticle has a large 
{\em electron} component on this site. Hence the 
electron  will have large  probability to tunnel  
into superconductor on this site and the tunneling 
intensity for electrons (positive bias) will be 
large. Conversely,  for those  sites the hole 
amplitude is small $|v(i)| \ll |u(i)|$ and the hole
intensity (negative bias) will be small. Similarly, 
for sites with large hole amplitudes $|v(i)| \gg |u(i)|$
the electron amplitude will be suppressed and this site
will be bright on the hole bias.  Therefore if there 
is a particular pattern for the large particle amplitude
(sampled on positive bias) on certain sites ${i}$, 
the complimentary pattern of bright sites for hole 
tunneling (on negative bias) will develop as a 
consequence of the inherent particle-hole mixture 
in superconductor. This is the main physics,  we believe, 
behind the cross rotation upon bias switch.

Our numerical results are summarized
on Fig. \ref{fig:best} where the particle and hole like intensity is plotted near the
impurity site.

To model the high-temperature superconductors we  
utilize the
highly-anisotropic structure of the cuprates and focus on a single layer of 
the material.  In the simplified model, the conduction electrons live on the 
copper sites, $i$, and can hop to the neighboring sites, $j$, with a 
certain 
probability measured by the quantity $t$.  In addition to that,
the electrons that occupy the neighboring sites feel mutual attraction of 
a strength $V$.  Formally, this model is represented by the Hamiltonian,
\beq
H_0 = -t \sum_{<i,j>, \sigma} {t c^{\dagger}_{i\sigma} c_{j\sigma}} 
	-V \sum_{<i,j>}{n_i n_j},
\eeq
were a quantum-mechanical operator $c^{\dagger}_{i\sigma}$ creates an electron on
site $i$, the operator $c_{j\sigma}$ eliminates an electron from the site $j$, and
$n_i = c^{\dagger}_{i\uparrow}c_{i\uparrow} + c^{\dagger}_{i\downarrow}c_{i\downarrow}$
 represents the electron density on site $i$.  The
electron spin, $\sigma$, can point up or down.  This model, 
referred to  as the $t-V$ model, is known to produce the d-wave pairing for
the electron densities close to one electron per lattice site.
The model, however, is invalid very close to the half-filled case (exactly one 
electron per Cu site) where the cuprates are no longer 
superconductors, but rather antiferromagnetic insulators.
For the model parameters we use $t = V = 300\ meV$.  
Such choice ensures that the electronic band structure of the cuprates is 
accurately represented, and the superconducting gap in the electronic density 
of states is on the order of $0.1 t = 30\ meV$.
The local impurity is introduced by modifying the electron energy on a particular
site.  The corresponding correction to the Hamiltonian is
\beq\label{eq:Himp}
H_{imp} = V_{imp}(n_{0\uparrow} + n_{0\downarrow}) - S_{imp}(n_{0\uparrow} - n_{0\downarrow}).
\eeq
The first term is the potential part of the impurity energy that couples 
to the total electronic density on site 0, and the second term describes
the magnetic interaction of the impurity spin and the electronic spin density
on the same site.  We assume that the impurity spin is large and can be treated 
classically, as if it were a local magnetic field.
We solve the impurity problem in the Hartree-Fock 
approximation, which replaces the two-body interaction in $H_0$ with an
effective singe-electron potential.  
Our goal is to determine $V_{imp}$ and $S_{imp}$ so as to match both the
location of the impurity states within the gap and the spatial distribution of 
their intensity. 

\begin{figure}[htbp]
  \begin{center}
    \includegraphics[angle = -90, width = 3.0 in]{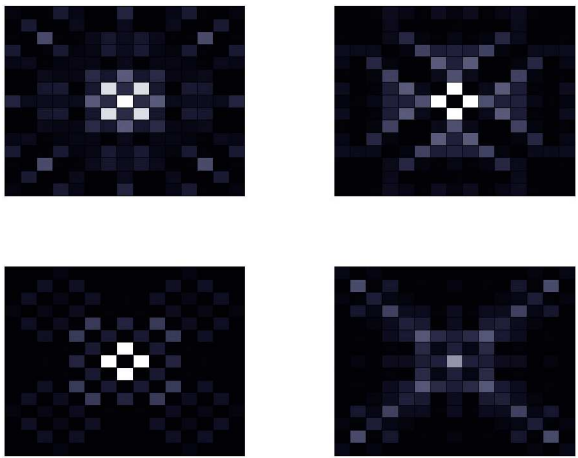}
	\vspace{0.5cm}

    \caption{Doping dependence of the impurity state intensity pattern. The impurity strength is 
	$V_{imp} = -3 t$.  Each square represents
	a lattice site.  The left-column patterns are for 
	the positive bias and the right column is for the negative bias.  The first row
	is the large doping case --- $20\%$ ($\omega_0 = 0.0661$); 
	the second row is the low doping case --- $7\%$ ($\omega_0 = -0.0347$).
	Notice how the positive and negative bias patterns get interchanged
	as a function of doping.}
    \label{fig:patts}
  \end{center}
\end{figure}

%%%%%%%% ADDED 2/14
The important aspect that we include in the treatment of the impurity problem is the
absence of the particle-hole symmetry.  The particle-hole symmetry is lost as soon as
we depart from the half-filled insulating part of the cuprate phase diagram, and hence is
related to the amount of doping.  While the asymmetry is not large, it results in 
two important effects.  First, it leads to redistribution of the spectral weight among the
impurity sites and its neighbors; second, it changes the position of the impurity level.
The first effect is closely analogous to the Friedel oscillations, which occur in the
vicinity of an impurity in a normal metal.  An impurity essentially plays a role of a
boundary condition imposed on the scattered electronic states.  Since the most important 
states are in the vicinity of the Fermi level, these states oscillate in space with the
corresponding filling-dependent Fermi wave vector.  This generates oscillation in the 
electronic density of states.  Similar phenomenon occurs for impurities in the 
cuprates.  However, it is compounded by the superconducting character of the quasiparticles, 
as well as the anisotropy imposed by the Cu-O lattice.  Figure \ref{fig:patts} demonstrates 
the high sensitivity of the impurity state intensity on the doping.

%%%%%%%%

\begin{figure}[htbp]
  \begin{center}
    \includegraphics[width = 3.0 in]{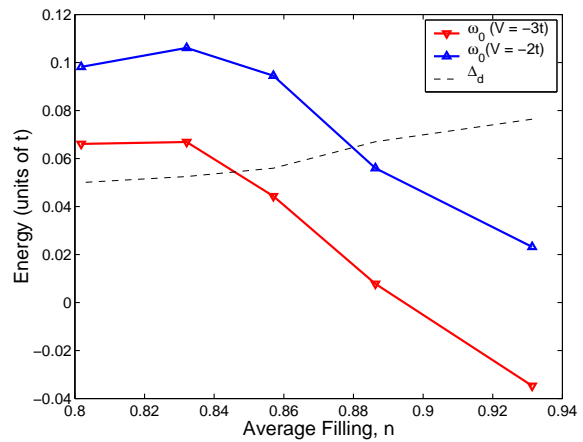}
	\vspace{0.5cm}

    \caption{Doping dependence of the impurity level position.  The red line is for 
	$V_{imp} = -3 t$; blue line is for $V_{imp} = -2 t$; the dashed line shows the level
	of the bulk d-wave gap obtained from the 16x16 lattice calculation 
	(the infinite 
	lattice result for the gap is about $50\%$ larger).}
    \label{fig:EnLev}
  \end{center}
\end{figure}

Similarly, the effect of changing the impurity level position by the particle-hole
 asymmetry (doping) has important consequences.  
In the particle-hole symmetric case, the impurity levels approach
the chemical potential as the strength of the impurity increases.\cite{balatsky}
In the limit of an
infinitely strong impurity (``unitary'' limit), the levels lie exactly at the chemical 
potential.  For a finite doping, the position of the levels changes. In figure \ref{fig:EnLev}
we show how the impurity level position, $\omega_0$, changes as a function of doping
for two different impurity strengths. 
Analytically, the 
change can be estimated from the non-self-consistent
$T$-matrix approximation.\cite{balatsky} One finds that for non-zero chemical potential 
$\mu$ (with $\mu = 0$ at the  half filling), the impurity levels are shifted by by an amount proportional to
$\mu/W$, where $W$ is the bandwidth.  Hence neglecting this effect can 
cause an error in the impurity strength
estimate.  In fact, in a doped superconductor the impurity levels can {\em cross} the
chemical potential at a finite impurity strength.  While this effect turns out to be 
not very important in the case of Ni impurity in BSCO, we believe that it is indeed 
relevant for Zn in BSCO.\cite{Pan1}  Unlike Ni, the Zn impurity levels appear to be very 
close to the chemical potential.  If we neglect the particle-hole asymmetry, this would
suggest that Zn impurity is in the unitary limit.  However, inclusion of the asymmetry shift 
would lead to a finite impurity strength, and would imply high sensitivity of the Zn level
position on the doping.
One of the characteristics of the unitary impurity states is that their spectral 
weight tends to zero on the impurity site, with the maxima positioned on the 
nearest neighbors.\cite{tsuchiura}  On the contrary, in the experiment, 
the spectral weight is maximized on the impurity site.  This suggests that neither Ni nor
Zn is in the unitary limit. More data on doping dependence of the position of
 Zn level inside the gap would help to clarify how relevant the particle-hole
  asymmetry effect is.

%%% END ADDED 2/14

%Role of the Bi layer filtering.  
%If we include the filtering then we can use V = -3 impurity, thus %making the impurity levels
%closer to the chemical potential (inside the gap).  If we use V = -2 %impurity, where 
%filtering is not necessary, then the impurity levels are outside the %gap for
%dopings above 12 percent.  What is the real doping for Ni in BSCO, %anyway?
%If we do not want to use filtering, we have to remove the claim that we %can explain position 
%of the level.
%So, I think it is better still to include the filtering.  What do you %think?
%lets wait for their reply about doping level . This part has top

\begin{figure}[htbp]
  \begin{center}
    \includegraphics[width = 3.0 in]{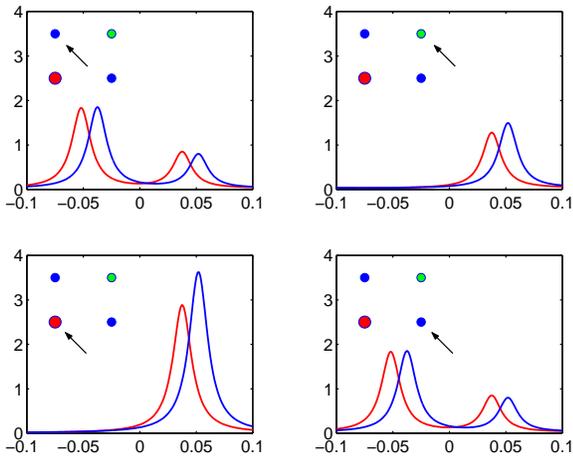}
	\vspace{0.5cm}

    \caption{Inclusion of a weak, $S_{imp} = 0.2 t$, spin component of the impurity potential
	causes the spin-degenerate
	spectral density peaks to split.  The red line is the spin-up branch and the blue is the
	spin down.  The bottom left chart is represents the spectral density at the impurity
	site; top-right --- on the next-nearest neighbor; top-left and bottom right --- nearest 
	neighbors. Due to the ferromagnetic coupling of the impurity spin and conduction
	electrons, the spin-up level lies lower than spin down.  The energy (horizontal axis) is 
	in the units of $t$.
	Impurity strength $V_{imp} = -3 t$, doping $14\%$. }
    \label{fig:SprSpin}
  \end{center}
\end{figure}

In our quantitative analysis we focus on the case of Ni impurity in BSCO.  The
 simple
impurity interaction model of Eq. (\ref{eq:Himp}) seems not to describe properly the Zn case. 
A possible reason
is that the effect of Zn is not fully local, and may include, for instance, modification of the 
hopping parameters in its vicinity and interactions with other bands, present in Cu-O plane.  Accounting for such effects would require a number of
extra fitting parameters, which would reduce credibility of the obtained results.  Hence, we 
restrict our attention to the Ni impurities, where the only fitting parameter needed is
the local impurity strength.  
We find that for an attractive impurity with a strength around $V_{imp} = - 3t = - 900\ meV$
 and the average 
fillings of about 0.85 electrons per site ($15\%$ doping), 
the impurity levels are situated within the gap with the
intensity distribution that corresponds to the experimental pattern 
around Ni impurity (figure \ref{fig:best}).
By including a weak spin part of the impurity interaction, $S_{imp} = 0.2 t = 
60\ meV$,
we reproduce the fine energy splitting of the impurity peaks also 
observed in the experiment of Pan et al. \cite{Pan1}.  The site-dependent spectral intensities
are shown in figure \ref{fig:SprSpin}.
 That the spin part should be present in
the interaction follows from the atomic structure of the Ni$^{2+}$  that substitutes for
Cu in the copper-oxygen layer and believed to have spin S = 1.  The strength of the conduction electron --
Ni spin coupling is extremely hard to determine from the first principles.  The 
close analysis of the Ni impurity states in the superconductor enabled us to 
extract the approximate strength of both the potential and the spin coupling 
between Ni impurity and electrons in BSCO.

Based on our theoretical analysis we can make the following predictions for the  properties of Ni 
impurities in BSCO: 
(1) For the lower-doped BSCO samples, the Ni-induced  peaks should shift closer to the chemical 
potential, and the patters should change according to Fig. \ref{fig:patts}; 
(2) In the presence of the in-plane magnetic field, there should be 
Zeeman splitting of the peaks on the scale of $0.1\ meV$/Tesla.  For some impurities, 
Zeeman splitting will enhance, and for the others suppress the intrinsic peak splitting 
due to the impurity spin (this effect depends on the relative alignment of the impurity 
spin and the external magnetic field); 
(3) for a similar potential strength impurity, but with a larger 
value of spin (Mn), the peak splitting in zero magnetic field should increase.

% ==============================

In conclusion, we find that a simple effective model well describes rich physics of the STM
images near Ni site. The model we adopted here is the simplest  
effective model of the real material, where only the 
on-site impurity effects are considered. As such, the model does not address
many aspects of the impurity influence on the electronic states of the
host material.   Surprisingly, even such a simple model exhibits a very rich set of
phenomena as a function of doping and impurity strength.
We find that to explain the experimental data we need to include both the 
non-magnetic scattering of carriers from Ni site as well as spin interaction between carriers
and the impurity spin. The most striking feature
observed in the experiment --- the rotation of the ``impurity cross'' as a function of 
bias --- appears to be a universal feature of the theoretical model.
This rotation is the 
manifestation of the quantum-mechanical nature of the quasiparticles in the 
superconducting state, and is a consequence of the unique particle-hole composition of the
quasiparticles.

%\vspace*{-2.2cm}
{99}

\end{multicols}


\begin{thebibliography}{99}

\bibitem{coral} H. C. Manoharan, C. P. Lutz, and D. M. Eigler, 
	Nature {\bf 403}, 512-515 (2000).


\bibitem{Pan1} S. H. Pan , 
   E. W. Hudson, K. M. Lang, H. Eisaki, S. Uchida, and J. C. Davis, Nature, {\bf 403}, 746-750 (2000)

\bibitem{Pan2} J. C. Davis, private communication.

\bibitem{Schrieff}   J.R.Schrieffer, {\it Theory of Superconductivity}, 
	Addison-Wesley, Redwood City, 1983.


\bibitem{shiba} H. Shiba, Prog. Theor. Phys. {\bf 40}, 435 (1968).

\bibitem{yulu} L. Yu, Acta Physica Sinica {\bf 21}, 75 (1965);

\bibitem{balatsky} A. V. Balatsky, M. I. Salkola, and A. Rosengren, 
Phys. Rev. B{\bf 51}, 15547 (1995);
A. V. Balatsky and M. I. Salkola, Phys. Rev Lett. {\bf 76}, 2386, (1996);  
M. Salkola, A. V. Balatsky, 
and J. R. Schrieffer, Phys. Rev. B {\bf 55}, 12648-12661 (1997).

\bibitem{byers} J. M. Byers, M. E. Flatte, and D. J. Scalapino, Phys. Rev. Lett. {\bf 71},
3363 (1993).

\bibitem{flattereview} M. E. Flatte and J. M. Byers,   Solid State Physics {\bf 52}
, 137, (1999). 

\bibitem{Yazdani} Yazdani, A., Jones, B. A., Lutz, C. P., Crommie, M. F. and Eigler,
   D. M. Science {\bf 275}, 1767-1770 (1997).



\bibitem{tsuchiura} H. Tsuchiura, Y. Tanaka, M. Ogata, and S. Kashiwaya, cond-mat/9911117 (1999).

%--------------------    
\end{thebibliography}
\end{document}